# Photoinduced Magnetic Force Microscopy: Enabling Direct and Exclusive Detection of Optical Magnetism

Jinwei Zeng[1,2], Mohammad Albooyeh[2], Mohsen Rajaei[2], Abid Anjum Sifat[2], Eric O. Potma[3]*, H. Kumar Wickramasinghe[2]*, Filippo Capolino[2]*.

**Affiliations:**

[1]Wuhan Nation Laboratory for Optoelectronics, Huazhong University of Science and Technology, Wuhan, Hubei, 430074, China

[2]Department of Electric Engineering and Computer Science, University of California Irvine, Irvine, CA, 92697, USA

[3]Department of Chemistry, University of California Irvine, Irvine, CA, 92697, USA

*Correspondence to:

Jinwei Zeng, zengjinwei@hust.edu.cn

Eric O. Potma, epotma@uci.edu

H. Kumar Wickramasinghe, hkwick@uci.edu

Filippo Capolino, f.capolino@uci.edu

**Abstract:** Modern optical nano-elements pursue ever-smaller sizes and individualized functionalities. Those elements that can efficiently manipulate the magnetic field of light boast promising future applications with a great challenge: the magnetic near field is irretrievable from conventional optical far-field characterization. Here we propose photoinduced *magnetic* force microscopy to directly and exclusively sense the magnetic field of light at the nanoscale. The proposed instrument exploits a magnetic nanoprobe with exclusive magnetic excitation under structured light illumination. The magnetic nanoprobe detects the photoinduced magnetic force, which is defined as the dipolar Lorentz force exerted on the photoinduced magnetic dipole in the nanoprobe. Since the resulting magnetic force is proportional to the incident magnetic field, the measured force reveals the magnetic near-field distribution at the nanoscale. The proposed instrument represents a fundamental step towards comprehensive electric and magnetic near-field detection and/or manipulation in single nano-element optical devices.

**One Sentence Summary:** Photoinduced magnetic force microscopy directly detects the magnetic field of light at nanoscale.

**Main Text**
Conventional optical devices usually detect and manipulate light in the far-field, i.e., at distances that are optically far from external and/or induced light sources. In the far-field, radiation is devoid of evanescent wave components and solely travels in the form of propagating waves. Consequently, optical sensing in the far-field lacks the subwavelength information that is present in the near-field of light-matter interaction, and is heavily influenced by non-local effects (*1–3*). In principle, far-field optical devices are diffraction-limited and are prone to environmental background noise (*3*). Therefore, measurements performed with far-field optical devices have intrinsically limited resolution and signal-to-noise ratio (SNR). The development of nanotechnology, however, forges new ways of light manipulation, where light-matter interactions are controlled and probed in the near-field (*3–8*). Near-field optical devices enable local manipulations at the subwavelength scale, free from interference with background radiation (*3*). Modern near-field optical nano devices commonly interact with the electric component of light (*9*), while the possibility to create optical magnetic nano devices provides tantalizing opportunities that are yet to be fully exploited. For example, the excitation and manipulation of the optical magnetic dipole in lanthanides can reach the molecular level (*4*, *10*), so that future magnetic storage device could be made from single lanthanide nano-structures with all-optical read/write capacity (*11*). In addition, the use of *optical magnetic force* in optical magnetic tweezers, magnetic levitations, etc. enables the mechanical control of nanoscale samples with superior accuracy in comparison to conventional mechanical technology (*12*, *13*). The pursuit of these promising opportunities demands an accurate and reliable method to characterize the magnetic component of near-field light with nanoscale accuracy. The method discussed here provides a way to exploit the elusive magnetic response of matter and to manipulate the usually concealed (i.e., masked) magnetic near-field (*14*, *15*). Special difficulties arise because the electric and magnetic fields are usually coupled with a ratio given by the local electromagnetic field impedance, and the magnetic response of common materials is overwhelmed by the electric response at optical frequencies (*16–20*). As a result, most detectors developed so far are based on matter interaction with the electric component of light.

To address this challenge, here we experimentally demonstrate a microscope device, called Photoinduced Magnetic Force Microscopy (PiMFM), which leverages the interaction of a magnetic dipole induced in a magnetic nanoprobe with the local magnetic component of light. We focus on the use of a silicon (Si) truncated cone as the magnetic nanoprobe, which features low dissipation losses at optical frequencies and benefits from practical fabrication protocols. The magnetic nanoprobe exhibits a degree of azimuthally geometrical symmetry that guarantees an exclusive optical magnetization and minimizes the electric dipole and quadrupole. We show that the resulting magnetization interacts with the magnetic component of light of a tightly focused azimuthally polarized beam (APB) (*21*). Importantly, we show that the magnetic nanoprobe provides an exclusive detection of the local *magnetic* field rather than the *electric* field, a property derived from the mutual symmetry between the nanoprobe structure and the illumination beam (*22*). With the aid of this probe, the PiMFM is sensitive to the photoinduced magnetic force, which is generated by the incident magnetic field. In this paper we demonstrate that PiMFM enables direct and exclusive registration of the magnetic near field of light at the nanoscale. We provide experimental evidence of exclusive magnetic force measurement at nanoscale, i.e., the force exerted on a magnetic dipole interacting with the magnetic component of light, as was theoretically predicted in (*23*, *24*).

**PiMFM Principle**

The PiMFM instrument is a dedicated tool for probing the photoinduced magnetic force, i.e., the force acting on a photoinduced magnetic dipole in a scanning nanoprobe, generated by the local magnetic component of light. The operating principle is inspired by the photoinduced (electric) force microscope (PiFM) which typically uses a gold-coated probe tip in a bottom- or side-illuminated atomic force microscopy (AFM) to accurately sense the electric near-field distribution (*3*, *21*, *25–27*). Unlike PiFM, the PiMFM replaces the gold-coated tip with a structure that exhibits a high magnetic polarizability at optical frequencies, rendering this opto-magnetic nanoprobe one of the key components of the PiMFM. We have designed a Si truncated cone structure that possesses a Mie-type magnetic resonance (*22*). A Mie-type magnetic resonance is achieved through assembling geometries of highly refractive dielectric nano-scatterers, which display circulating electric displacement currents that are associated with a magnetic dipole moment (*28*, *29*). We choose silicon as the tip material because of the low dissipative loss of Si and the ease of fabrication afforded by the use of commercially available Si-probe tips as a template.

Another difference between PiMFM and PiFM is the use of an azimuthally polarized excitation beam. Under an APB illumination, the magnetic mode of the Si truncated cone is exclusively excited (*22–24*). Since the APB possesses an azimuthally symmetric electric field distribution that circulates about its beam axis and fully vanishes on-axis, it exhibits an axial (longitudinal) magnetic field (*30*, *31*). The magnetic field component of the APB induces the signature circulating displacement current inside the Si truncated cone. When the cone is properly aligned with the beam axis, this geometry provides the exclusive Mie-type magnetic resonance excitation and forms the photoinduced magnetic dipole oriented along the longitudinal direction, i.e., collinear with the beam propagation axis (*22*). In the case when the truncated cone and the incident beam are slightly misaligned, we show that the magnetic response of the nanoprobe is still dominant compared to the electric response, underlining the robustness of PiMFM to setup tolerances (*22*).

The concept of the PiMFM is shown in Fig. 1. The incident APB is tightly focused by the bottom oil-immersion objective lens through a glass coverslip onto the Si truncated cone (i.e., the magnetic nanoprobe) as indicated by Fig. 1 (a). The incident APB excites the magnetic mode in the Si probe, inducing a magnetic dipole. When the nanoprobe of the PiMFM approaches the glass slip (or substrate) to within a few nanometers' distance, the interaction between the nanoprobe and the substrate under APB illumination is modeled with two interacting magnetic dipoles: the induced magnetic dipole in the nanoprobe and its image, which accounts for the substrate effect, as indicated by Fig. 1 (b). Such a system of two closely spaced magnetic dipoles is shown in Fig. 1 (c), and generates a magnetic dipole-dipole interaction force, that is measured by the PiMFM.

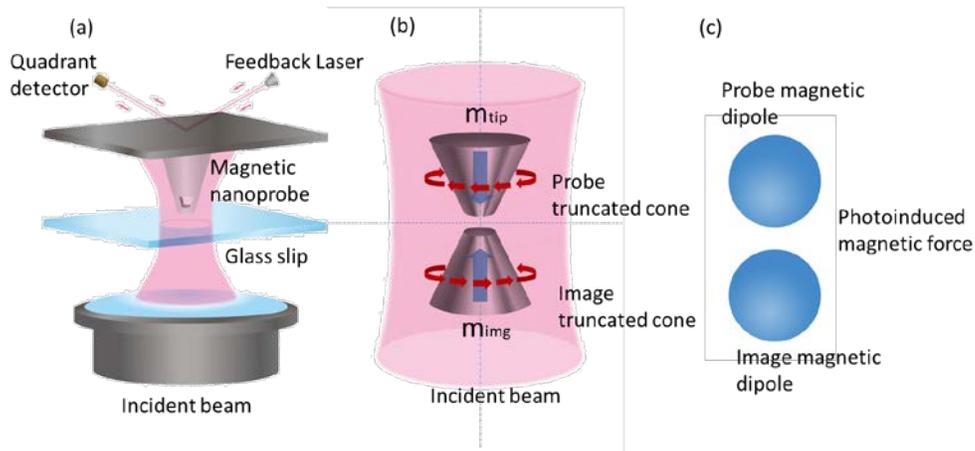

Fig. 1. (a) the schematic of the PiMFM. (b) the zoom-in region of the Si truncated cone working as photoinduced magnetic nanoprobe and its image in the glass slip (or substrate). The rotating arrows represent the excited electric field at the magnetic resonance condition and the bold blue arrows show the directions of probe and image magnetic dipoles $m_{tip}$ and $m_{img}$, respectively. (c) two photoinduced magnetic polarized nanoparticles exerting a magnetic force on each other.

**From Magnetic Field to Magnetic Force**

One essential task of the proposed PiMFM is to relate the measured total optical force to the *incident* magnetic field. For this purpose, we analyze the physical origin of the optical force exerted on a nanoprobe in force microscopy. In general, the optical force on an object is due to the Lorentz force density, provided by the distribution of the induced charge and current densities $\rho_{ind}(\mathbf{r},t)$ and $\mathbf{J}_{ind}(\mathbf{r},t)$, respectively. The Lorentz force *density* $f(\mathbf{r},t)$ exerted at any position $\mathbf{r}$ of an object, at time $t$, subject to electromagnetic field with electric field $E(\mathbf{r},t)$ and magnetic induction $B(\mathbf{r},t)$ reads $f(\mathbf{r},t) = \rho_{ind}E + J_{ind} \times B$. Here time dependent quantities are in italic. Therefore, the *total* force on an object with volume $V$ is given by $F_{tot}^{Lorentz}(t) = \int_V f\, dv$. Considering time-harmonic fields with $e^{-i\omega t}$ dependence, using phasors (letters with a normal font) the time-average total optical force is given by $\mathbf{F}_{tot}^{Lorentz} = \frac{1}{2}\text{Re}\left(\int_V \rho_{ind}\mathbf{E}^* + \mathbf{J}_{ind} \times \mathbf{B}^* dv\right)$.

Next we assume that the nanoprobe response is well approximated by its electric and magnetic dipolar responses, hence we neglect the contributions from higher order multipoles. The proof for the validity of this assumption is given in the supplementary materials. Therefore, the time-average total optical force reduces to (*13*, *32*):

$$\mathbf{F}_{tot}^{dipole} = \frac{1}{2}\text{Re}\left\{\left(\nabla\mathbf{E}^{loc}(\mathbf{r})\right)^* \cdot \mathbf{p} + \mu_0\left(\nabla\mathbf{H}^{loc}(\mathbf{r})\right)^* \cdot \mathbf{m} - \mu_0\frac{ck^4}{6\pi}(\mathbf{p}\times\mathbf{m}^*)\right\}. \quad (1)$$

Here, $\nabla\mathbf{E}^{loc}$ and $\nabla\mathbf{H}^{loc}$ are the gradient of the local electric and magnetic fields, respectively; $\mathbf{p}$ and $\mathbf{m}$ denote the electric and magnetic dipole moments in the nanoprobe, respectively (see Eq. (S5)); $c$ is the speed of light, $k$ is the free-space wave-number, $\mu_0$ is the free-space permeability; and the superscript * denotes complex conjugation. When the nanoprobe is a photoinduced dominant electric or magnetic dipole (with one dominating on the other), the last term $\mu_0 ck^4(\mathbf{p}\times\mathbf{m}^*)/(6\pi)$ in Eq. (1) is negligible. This yields $\mathbf{F} \approx \mathbf{F}_p^{dipole} + \mathbf{F}_m^{dipole}$, where the corresponding photoinduced electric and magnetic dipolar force expressions are:

$$\mathbf{F}_p^{dipole} = \frac{1}{2}\text{Re}\left\{\left(\nabla\mathbf{E}^{loc}(\mathbf{r})\right)^* \cdot \mathbf{p}\right\}, \quad (2)$$

$$\mathbf{F}_m^{dipole} = \frac{1}{2}\mu_o\text{Re}\left\{\left(\nabla\mathbf{H}^{loc}(\mathbf{r})\right)^* \cdot \mathbf{m}\right\}, \quad (3)$$

respectively. We recall that the PiFM uses a gold-coated probe tip which commonly detects the *electric* dipolar force exerted on an electric dipole (the nanoprobe of the PiFM) resulting from two closely spaced interacting photoinduced electric dipoles given by Eq. (2) (*21*). Therefore, in the PiFM system, as proven in (*21*), the longitudinal electric dipolar force is related to the *incident* transversal electric field $E_t$ as $F_{p,z}^{\text{dipole}} \propto |\alpha^e E_t|^2$, where $\alpha^e$ is the electric dipolar polarizability of the nano-probe. As a fundamental conclusion, in a PiFM system the electric dipolar force is directed along the beam axis and it is directly proportional to the incident electric field intensity via the proportionality constant of the electric polarizability of the nano-probe.

Instead, the basic principle of the PiMFM is to relate the incident magnetic field (oscillating at optical frequency) to the measured longitudinal force. To understand this principle, let us consider the time-averaged magnetic dipolar force $\mathbf{F}_m^{\text{dipole}}$ given by Eq. (3). We recall that the local magnetic field in Eq. (3) acting on the nanoprobe is composed of two field terms: one is the incident field and the other one is the field generated by the presence of the substrate. However, the optical magnetic force should provide useful information about the excitation field only, therefore we equivalently represent the effect of interaction fields into the incident field following the derivation in the Supplementary Materials. As a result, the longitudinal magnetic dipolar force acting on the magnetic nanoprobe is related to the *incident* longitudinal magnetic field $H_z^{\text{inc}}$ as

$$F_{m,z}^{\text{dipole}} \propto |\alpha_{\text{tip}}^m H_z^{\text{inc}}|^2, \qquad (4)$$

where $\alpha_{\text{tip}}^m$ is the dipolar magnetic polarizability of the magnetic nanoprobe.

In summary, Eq. (4) shows that the magnetic dipolar force along the longitudinal beam direction is proportional to the longitudinal incident magnetic field intensity via the magnetic polarizability of the magnetic nanoprobe. In accordance of the linearly proportional relationship between the electric and magnetic dipolar forces and the incident field, hereby we name the electric dipolar force and magnetic dipolar force as "electric force" and "magnetic force", respectively.

Next, we demonstrate the overall dominant contribution of the magnetic dipolar force to the total optical force exerted on the magnetic nanoprobe. To address this point, we have included a rigorous analytical and numerical derivation in the Supplementary Materials. In conclusion, we demonstrate that a dielectric Mie scatterer can be reasonably considered as a dominant magnetic dipole when the magnetic resonance is excited by an incident APB. Most importantly, we show that at the magnetic resonance (that we refer to as "on-state") the Si truncated cone acts as a magnetic nanoprobe, while scanning within the proximity of the incident APB. Importantly, we demonstrate that the longitudinal component of the total, time-averaged optical force $F_{\text{tot},z}^{\text{Lorentz}}$ exerted on the scanning nanoprobe is almost entirely contributed by the longitudinal component of the time-averaged magnetic dipolar force defined as

$$F_{m,z}^{\text{dipole}} = \tfrac{1}{2} \mu_o \operatorname{Re}\left( \left(\nabla \mathbf{H}^{\text{loc}}(\mathbf{r})\right)^* \cdot \mathbf{m} \right)_z, \qquad (5)$$

i.e., we state that $F_{tot,z}^{Lorentz} \approx F_{m,z}^{dipole}$. For the purpose of easy representation, we study a system of two closely spaced spheres as two Mie scatterers shown in Fig. 1a, which represents a simple physical model for the interactive system of the nanoprobe and the substrate. In Fig. 2 we show the total optical force $F_{tot,z}^{Lorentz}$ exerted on the nanoprobe sphere when the two spheres are displaced from the APB axis (the derivation details are appended in the Supplementary Materials). We have also included three more curves in the Figure, representing $F_{B,z}^{Lorentz} = \frac{1}{2} \text{Re} \left( \int_V \mathbf{J}_{ind} \times \mathbf{B}^* dv \right)_z$, $F_{p,z}^{dipole} = \frac{1}{2} \text{Re} \left( \left( \nabla \mathbf{E}^{loc}(\mathbf{r}) \right)^* \cdot \mathbf{p} \right)_z$ and $F_{m,z}^{dipole}$ to corroborate our most important conclusion.

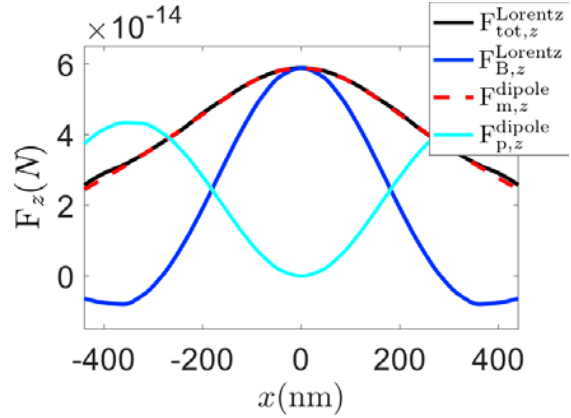

Fig. 2. Comparison of time-averaged optical forces exerted on the spherical Si probe, in the system of two closely spaced probes shown in Fig. 1 (a), under APB illumination from the bottom. The abscissa represents the displacement of both spheres from the APB axis. The optical wavelength is 610 nm, corresponding to the magnetic Mie resonance. The curves represent results evaluated with the accurate total Lorentz force density in Eq. (S1) and the dipole approximation formulation in Eqs. (2)-(3). See also Eq. (S3). This result shows that $\mathbf{F}_{tot,z}^{Lorentz} \approx F_{m,z}^{dipole}$, for any displacement. See Supplemental Materials for the incident beam as well as the geometrical parameters.

The main observation is that the total optical Lorentz force $\mathbf{F}_{tot,z}^{Lorentz}$ is equal to the magnetic dipolar force $F_{m,z}^{dipole}$ with very good accuracy. Note that the contribution of the $F_{B,z}^{Lorentz}$ term, the $\int_V \mathbf{J}_{ind} \times \mathbf{B}^*$ term in the Lorentz formula, which includes the effect of the current density $\mathbf{J}_{ind}$ that creates the artificial magnetization in the Mie resonator, does *not* represent the main contribution to the total optical force. Also, the term $F_{p,z}^{dipole}$ does not correlate with the observed total optical force. Therefore, we conclude that the photoinduced magnetic dipolar force $F_{m,z}^{dipole}$ is the main contributor to the total optical force $F_{tot,z}^{Lorentz}$.

Next, as a comparison we investigate the force on a photoinduced magnetic nanoprobe when the operating optical wavelength is off the magnetic resonance (referred to as the "off-state") to show the importance of a proper selection of optical frequency in the design. As shown in the Supplementary Materials (see Fig. S4), when the nanoprobe is nonmagnetic, i.e., is in the off-state, the total optical force exerted on the dielectric nanoprobe includes nontrivial contributions from both electric and magnetic dipolar forces, and hence, the Lorentz force can neither be attributed to the magnetic nor electric dipoles.

As a result, using a probe in the on-state, the PiMFM can map the incident magnetic field with super-resolution via the overall optical force detection mechanism. The azimuthally symmetric field distribution of an APB makes this beam an ideal choice for illumination in the proposed PiMFM system. Under APB illumination, we have theoretically showed that a magnetically polarizable nanoprobe can detect the photoinduced magnetic force of light, hence detecting the optical magnetic near field. This result is in analogy to what was achieved in a previous work (*21*) that demonstrated that an electric nanoprobe such as a gold-coated tip can detect the photoinduced electric force and hence detect the electric near field (*21*). As a result, when an electric or magnetic nanoprobe raster-scan a tightly focused APB, the measured force maps shall exhibit the signature APB electric and magnetic profiles, i.e., donut and solid center circular shapes, respectively. For the purpose of comparison and demonstration, we design the on- and off-state Si truncated cone samples shown in Fig. 2(a), and we numerically analyze these states with the finite element method (*33*) implemented in the commercial simulation software COMSOL Multiphysics (*34*).

Considering the chosen laser wavelength of 670 nm, the truncated Si cone is at its on-state and at its off-state when its height is equal to 90 nm and 140 nm, respectively, while in both cases the bottom (shorter) diameter is 150 nm and side tilt angle is 20°. As shown in Fig 3(b), the simulated magnetic force spectrum for the truncated cone placed 5 nm above the top of the glass substrate shows the unique strong force feature of on-state conditions at 670 nm. The off-state probe exhibits a much smaller magnetic dipolar force at 670 nm, and indeed it exhibits a peak at a longer wavelength of 810 nm. The magnetic-dipolar optical force exerted on the Si truncated cones in Fig. 3(a) has been evaluated for the truncated Si nanoprobes 5 nm above the glass substrate using Eq. (5). In the calculations we have assumed the incident APB light is focused to the top surface of the substrate. The incident APB has an azimuthal electric field distribution such that (*21, 31*) $\mathbf{E} = \hat{\boldsymbol{\varphi}}(V/\sqrt{\pi})(2\rho/w^2)e^{-\zeta(\rho/w)^2}e^{-2i\tan^{-1}(z/z_R)}e^{ikz}$ where $w = w_0\sqrt{1+(z/z_R)^2}$, $\zeta = 1 - iz/z_R$, $z_R = \pi w_0^2/\lambda$, $\rho$ is the radial distance from the beam axis, and $w_0$ is the minimum beam waist which is chosen to be $0.7\lambda$ with $\lambda$ being the operating wavelength. Moreover, the amplitude $V$ is determined from the incident power $P_{\text{inc}} = |V|^2 \left[1 - 1/\left(2(\pi w_0/\lambda)^2\right)\right]/(2\eta_0)$, where $\eta_0$ is the free space wave impedance. The incident power is assumed to be equal to 150 µW. The longitudinal component of the magnetic field $H_z$ is derived from the expression of the electric field using the Maxwell-Faraday equation $\mathbf{H} = \nabla \times \mathbf{E}/(i\omega\mu_0)$ leading to $H_z = A_0\left[1 - (\rho/w)^2 \zeta\right] e^{-\zeta(\rho/w)^2} e^{-2i\tan^{-1}(z/z_R)} e^{ikz}$ with $A_0 = -(V/\sqrt{\pi})\left[4i/(w^2\omega\mu_0)\right]$. The expressions for the azimuthal electric and longitudinal magnetic fields of an APB clearly demonstrate a vanishing value of the electric field and a non-vanishing value of the magnetic field at the beam axis.

In Fig. 3(c, d) we show the total longitudinal force map relative to the nanoprobe (i.e., the truncated Si cone) lateral scan, perpendicular to the APB axis at 670 nm wavelength. The time-averaged longitudinal force on the truncated Si cone over the substrate has been numerically computed using the *z* component of the total optical force in Eq. (1). As illustrated in Fig. 3(c, d), the simulated force profiles for the on-state and off-state nanoprobes exhibit a bright center circular spot and a donut shape, respectively, in accordance with the magnetic and electric field profiles of the APB, respectively.

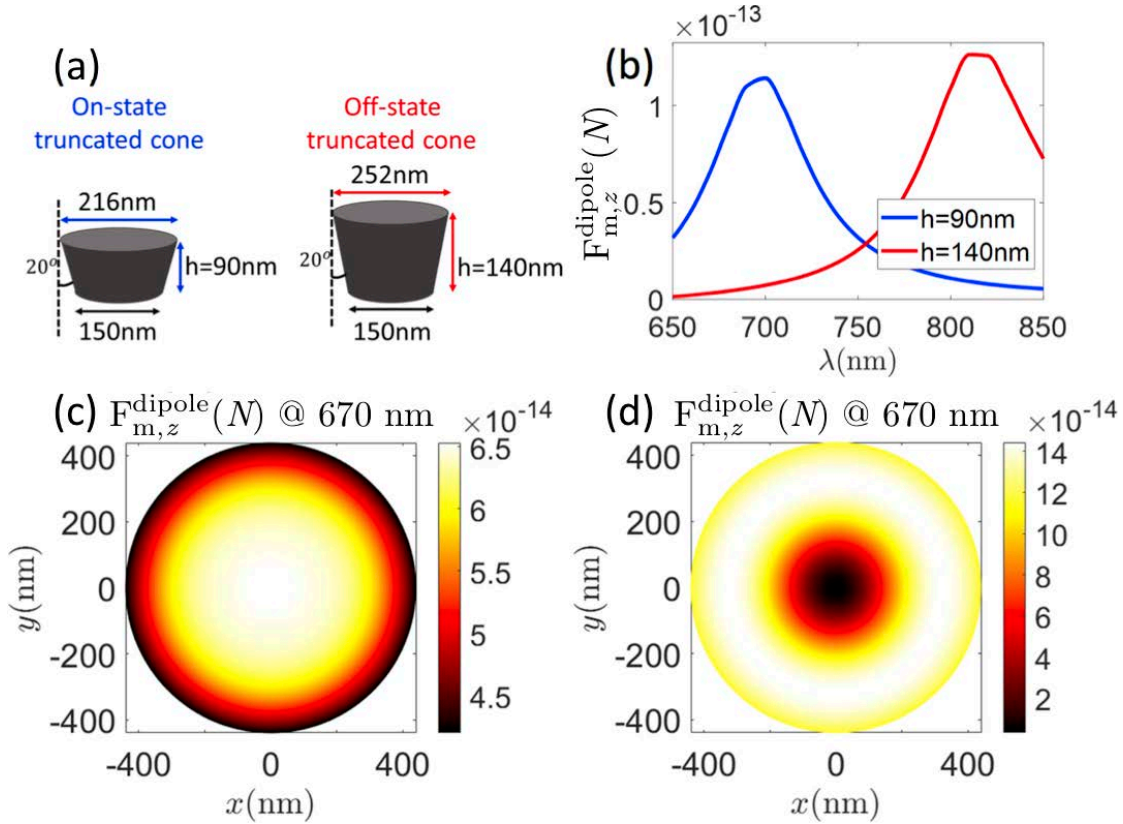

Fig. 3. (a) Designed on-state (left) and off-state (right), when operating at 670 nm, Si truncated cones with height 90 nm and 140 nm, respectively. The side tilt angle and bottom diameter for both structures are 20° and 150 nm. (b) Simulated magnetic force spectrum of the on-state probe (blue) and off-state probe (red), 5 nm above a glass substrate, under APB illumination. The on-state probe generates much larger magnetic force compared to off-state probe at 670 nm. (c, d) Simulated force profile in the transverse (x-y) plane for the on-state probe and off-state probe, respectively at 670 nm; The simulations have been performed with a focused APB beam of incident power of 150 µW with minimum beam waist parameter $w_0 = 0.7\lambda$.

**Experimental Demonstration**

Based on the design of the on-state and off-state nanoprobes made of truncated Si cones, we accordingly fabricated them, and we also use a blunt Si tip and a sharp Si tip as control groups. The incident APB has the wavelength of 670 nm and is focused by a bottom-illuminated oil immersion lens through the glass slip (substrate) onto the nanoprobes, as in Fig.1. The incident power into the PiMFM system is about 150 µW. The tightly focused APB on the front surface of the glass coverslip has a diameter of approximately 500 nm. Further details on fabrication, the optical setup and characterization can be found in the Supplementary Materials.

Figure 4 shows the experimental results: the on-state nanoprobe (the one that is experiencing a dominant photoinduced magnetic dipole at 670 nm) produces a solid-center circular spot force map. The off-state probe produces a donut-type circular-spot force map, whereas the blunt Si probe tip produces a clear donut-shape circular spot force map, and the unmodified sharp Si probe tip produces no detectable force map.

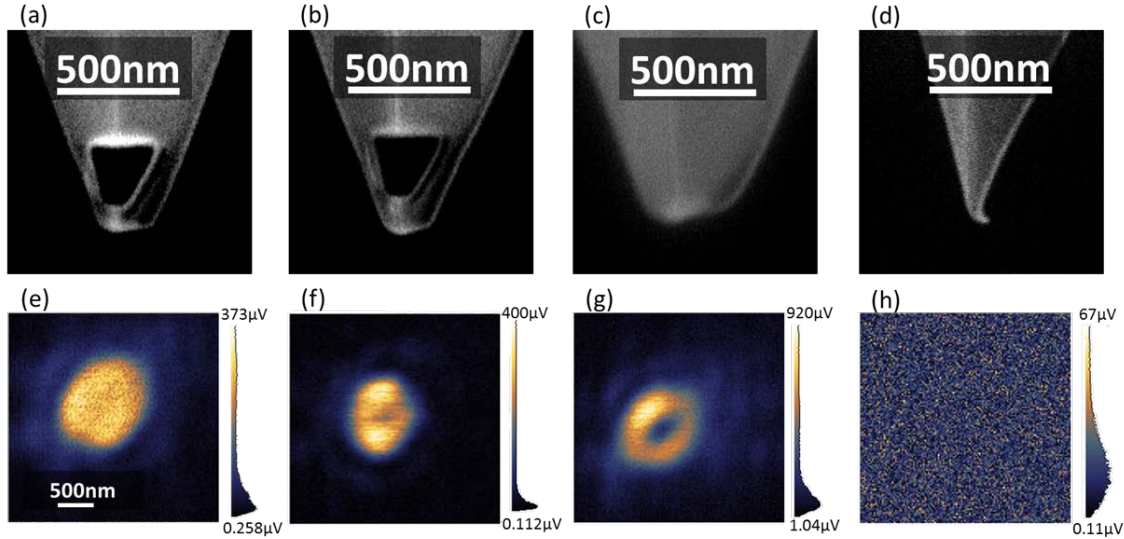

Fig. 4. (a-d) Focused ion beam images of the "on-state" Si truncated cone probe, "off-state" Si truncated cone probe, blunt Si probe, and sharp Si probe, respectively; (e-h) Corresponding measured force maps upon APB illumination from the bottom of the glass slip using the on-state Si truncated cone probe, off-state Si truncated cone probe, blunt Si probe, and sharp Si probe, respectively. The on-state probe (a) is measuring the solid-center circular spot (e) typical of the APB magnetic field.

Details on the physical properties of the blunt and sharp Si tip control groups in Figs. 4(c) and 4(d) are reported in the Supplementary Materials. Here we summarize that without a proper nanostructure at the tip end to support optical magnetic resonance, both the blunt and sharp Si tips exhibit a very weak magnetic response and cannot detect the magnetic force of light. In addition, for the blunt tip, the effective tip apex resembles a truncated cone structure with a substantial width, where an effective electric dipole is created inside the structure and hence detects the electric force of light. Therefore the force map assumes the donut shape in Fig. 4(g), as already discussed in earlier experiments based on electric dipole-electric dipole interaction(*21, 22, 35*). A similar mechanism is also present for the sharp tip, however, in this case the effective electric dipole is created further away from the tip and hence the distance between the effective electric dipole to the glass slip (i.e., to the image dipole) is much larger than that in the blunt tip case. We recall that the optical electric force is proportional to $1/d^4$, where $d$ is the dipole-dipole distance, and that the tip and its image in the substrate make the two dipoles. The reader may refer to the Supplementary Materials for a detailed discussion about the two equivalent scenarios: an illuminated electric dipole with a substrate system and an illuminated dipole with its image dipole (to replace the substrate) system. In such a modeling the longitudinal component of the time-averaged electric dipolar force is approximated as (*21*): $F_{p,z}^{\text{dipole}} \approx -3\Re\{\alpha_{\text{tip}}\}\Re\{\alpha_{\text{img}}\}|E_{\text{inc}}|^2/(8\pi\varepsilon_o d^4)$ where $\alpha$ is the electric polarizability of the tip/image with a unit of [$Fm^2$], $d$ is the dipole-dipole distance, and $E_{\text{inc}}$ is the transverse component of the incident electric field at the position of the tip dipole. Therefore, the overall optical force exerted on the sharp Si tip is undetectable due to the large distance between the tip and its image dipoles as shown in Fig. 4(h), while it can be detected when using a blunt tip over a substrate because of a shorter $d$.

The comparison in the measured force maps from different probes in Fig. 4 provides an important demonstration for the proposed magnetic field detection mechanism. From the on-state nanoprobe to the off-state nanoprobe to the blunt tip nanoprobe, the measured force maps transition from the solid-center circular shape to the donut shape. This observation is in accordance with the theoretical analysis that the photoinduced magnetic dipole contribution gradually decreases in strength for these nanoprobes, respectively. As a result, the center area of the force map given by 4(e) represents the target magnetic field intensity profile of the incident APB.

**Discussion**

Here we provide an overall analysis and evaluation of the results. First, we observe that the measured force maps in Fig. 4(e-f) exhibit some asymmetry. According to a previous study of a related phenomenon (*21*), this asymmetry is caused predominantly by the anisotropy of the probe (*21*). Since the original commercial Si tip has ridges, it is naturally anisotropic. Also, the Focused Ion Beam (FIB) fabrication of the tip will induce an arbitrary deformation, thus inevitably making it anisotropic. We expect that using a perfectly symmetric nanoprobe would largely improve the symmetry of the measurement, as demonstrated by previous research which indicated that the use of a spherical gold particle can remedy the measured asymmetry by a gold-coated tip (*35*).

Second, we discuss the signal to noise ratio (SNR) and measurement resolution. Here we define the SNR of the force map as the maximum force amplitude divided by the average background amplitude (the force amplitude is represented in microvolts measured by the quadrant photo-detectors of the PiMFM as shown in the scale bars). Therefore, Fig. 4(e-g) gives the SNR of 18.6, 24.4, and 14.4, respectively. Especially, we define the "magnetic SNR" as the force amplitude at the center of the force profile in Fig. 4 (e) divided by the background noise, which gives the magnetic SNR of about 15. We note that the incident power into the microscope is around 150 µW, and we observe the remarkable fact that the measured magnetic SNR is several times higher than in many conventional NSOM or optical force measurements of structured light at the subwavelength scale (*20*, *36*, *37*). The electric force map of structured light detected by a gold-coated probe in PiFM can have a SNR in excess of 100 (*22*), which is supported by strong plasmonic resonances and thus exhibits a high field-to-force coefficient. Nevertheless, the measured magnetic SNR achieved by the proposed PiMFM is a milestone in the direct detection of magnetic field of light at nanoscale.

Since the incident APB has no significant subwavelength structure, it is not suitable for defining the measurement resolution through the measured force map. However, following the basic concepts of force microscopy, resolution is determined by the size of nanoprobe (*22*, *25*, *38*). As a result, we can estimate the resolution in the force map shown in Fig. 3 (e-f) to be approximately the size of the probe-apex as shown in Fig. 4(a-c), i.e. about 200 nm.

The size of the photoinduced magnetic nanoprobe depends on the wavelength of the chosen magnetic resonance. Depending on the potential improvement of design and fabrication techniques, it is possible to make an effectively smaller nanoprobe to replace the current truncated cone structure. Based on Mie scattering principles, a Si magnetic Mie scatterer may have the size of about 1/4 of the functional wavelength. Furthermore, considering the bandwidth of the magnetic resonance as shown in Fig. 3 (b), the minimum size of the Mie scatterer can reach sub-100 nm for it to span the visible range of the spectrum, underlining that higher resolution can be achieved.

Finally, we demonstrated the performance of PiMFM using a Si truncated cone as the nanoprobe tip and were able to directly detect the local magnetic field using an APB illumination. We acknowledge that sensing of the magnetic force can also be achieved through the interaction between a Si tip and a special magnetic resonator. The effective magnetic dipole enabled by the resonator can induce an image dipole in the tip, and thus detect the magnetic force of light(*20*). However, we emphasize that this method, due to the lack of azimuthal symmetry, cannot completely suppress the electric response of the probe and thus enable the exclusive detection of the magnetic force. Our proposed PiMFM is unique in exclusive magnetic force detection. Moreover, the demonstrated concept is far more general and in principle can be applied to any optical field structure. This requires a nanoprobe tip that has an exclusive magnetic response to any optical beam. One possibility is to use a nanoprobe tip composed of rare-earth elements with a clear footprint of magnetic dipolar atomic transitions as demonstrated in Ref. (*10*, *42*).

**Conclusion**

We have conceived and experimentally demonstrated a PiMFM instrument that is able to directly and selectively detect the magnetic field of light at the nanoscale. We have employed a Si truncated cone as the photoinduced magnetic nanoprobe mounted in a force microscope with axis-aligned APB illumination. The mutual azimuthal symmetry of the incident APB and the nanoprobe enables exclusive magnetic excitation of the photoinduced nanoprobe; we have demonstrated that the total optical force exerted on the PIMFM nanoprobe is contributed predominantly by the magnetic dipolar force. The measured photoinduced magnetic force is proportional to the local magnetic field detected by a scanning nanoprobe near the center of the focused APB. As a result, the PiMFM instrument was able to detect the magnetic field intensity profile of an incident APB. The functionality and idea of PiMFM may inspire emerging research areas in nanophotonics like the characterization and exploitation of elusive material-magnetic properties in nano-structures, which may fundamentally revolutionize the multidisciplinary fields of bio/chemical sensing, quantum information processing, opto-mechanical manipulation, etc.

# Supplementary Materials for

Photoinduced Magnetic Force Microscopy: Enabling Direct and Exclusive Detection of Optical Magnetism

Jinwei Zeng[1,2], Mohammad Albooyeh[2], Mohsen Rajaei[2], Abid Anjum Sifat[2], Eric Potma[3], H. Kumar Wickramasinghe[2], Filippo Capolino[2].

[1]Wuhan National Laboratory for Optoelectronics, School of Optical and Electronic Information, Huazhong University of Science and Technology, Wuhan 430074, China
[2]Department of Electric Engineering and Computer Science, University of California Irvine, Irvine, CA, 92697, USA
[3]Department of Chemistry, University of California Irvine, Irvine, CA, 92697, USA

correspondence to:
Jinwei Zeng, zengjinwei@hust.edu.cn
Eric O. Potma, epotma@uci.edu
H. Kumar Wickramasinghe, hkwick@uci.edu
Filippo Capolino, f.capolino@uci.edu

**Materials and Methods**

**Si probe on top of glass substrate illuminated by an APB**

As the essential part of the proposed photoinduced magnetic force microscopy (PiMFM) system, in this subsection we discuss a physical modeling for the Si probe on top of the glass substrate illuminated by an incident APB.

As has already been discussed in a previous work (*22*), a subwavelength Si nanoparticle can be considered as a pure magnetic dipole scatterer provided that a proper symmetric excitation is applied to exclusively excite its magnetic dipole moment. In that work, we have experimentally obtained the force map, based on the electric field component of light, exerted on a gold tip on top of a Si truncated cone located on a glass substrate and illuminated by an APB. Therefore in that work we measured the electric dipolar force exerted on the scanning nanotip (that is an electric nanoprobe). We have shown that the measured force is proportional to the incident electric field intensity, and that is the reason we call it the electric force of light (*21*). Based on experimental results and on rigorous theoretical analysis based on the calculation of multipole moments, in (*22*) we have *indirectly* demonstrated (from the measurement of the electric force) that a silicon disk is an effective magnetic scatterer at a specific wavelength range and under APB excitation.

Here we use the magnetic scatterer investigated in (*22*) as a probe of the magnetic force, and devise a new force microscopy technique which enables *direct and exclusive* acquisition of the optical force due to the magnetic field component of light rather than its electric counterpart. As noted in the main body of the manuscript, we call this technique Photoinduced Magnetic Force Microscopy (PiMFM) since it enables the direct measurement of magnetic field component of light, in contrast of the PiFM system that detects the electric field component of light.

Considering the dipole approximation of the Si magnetic nanoprobe of the PiMFM shown in Fig. S1, the validity of which is provided in the following, the simplest scenario that mimics the interaction between the Si probe (as a dipole scatterer) and the glass substrate under APB illumination is shown in Fig. S1 (Right). First we replace the Si magnetic nanoprobe with a Si sphere that carries a Mie "magnetic" resonance, then we replace the overall effect of the substrate with an image dipole. As a result, we consider two closely spaced Si sphere that mainly provide two coupled magnetic dipoles; one representing the probe whereas the other illustrating its image. In the following we also explore the role of the other contributions like the electric dipole and quadrupoles, demonstrating the dominance of the magnetic dipole in generating the total force on the nanoprobe. Fig. S1 illustrates these two scenarios, i.e., the real scenario of the Si probe on top of the glass substrate, and the equivalent scenario of the two identical spheres illuminated by an APB from the bottom (*43*).

The equivalent problem of two interacting magnetic dipolar dipoles is treated analytically and it provides a basic physical insight of the real scenario. During the whole analysis, we keep in mind that the main goal is to find the optical force exerted on the Si probe on top of the substrate when illuminated by an APB from the bottom (see Fig. S1).

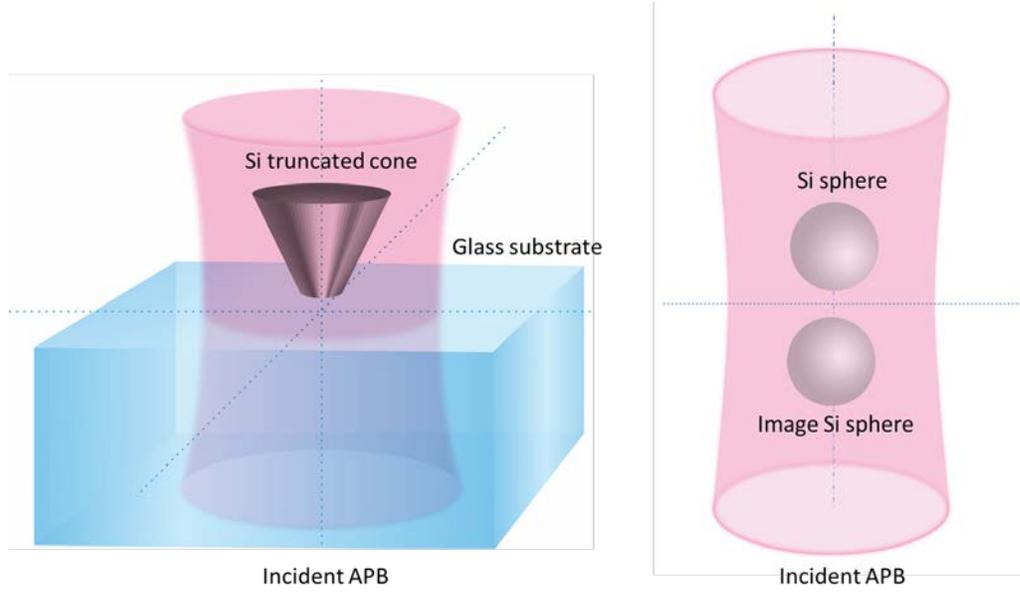

**Fig. S1.** Real scenario [left] and its physical modeling [right]. The Si probe on the top of a glass substrate is illuminated with an APB incident from the bottom side. In the physical modeling, the truncated cone shape of the probe is modeled as a sphere shape for the convenience of analytical derivation. And the substrate is replaced by the image sphere to mimic the electromagnetic interaction between the probe and the substrate.

### The Lorentz force density and the total optical force

The optical force on an object is derived from the Lorentz force density generated by a distribution of charge and current densities $\rho(\mathbf{r},t)$ and $J(\mathbf{r},t)$, respectively. The Lorentz force *density* $f(\mathbf{r},t)$ exerted at any position $\mathbf{r}$ is given by

$$f(\mathbf{r},t) = \rho E + J \times B. \tag{S1}$$

where $E(\mathbf{r},t)$ is the electric field vector, and $B(\mathbf{r},t)$ is the magnetic induction vector. Here all italic symbols define position and time dependent quantities. Therefore, the *total* force on the proposed object with volume $V$ containing these charge and current distributions reads $\mathbf{F}_{\text{tot}}^{\text{Lorentz}}(t) = \int_V f\, dv$. Since we do not consider any "impressed" (or forced) source within the magnetic nanoprobe volume in our scenario, then, the force is calculated using the induced charge and current densities, $\rho_{\text{ind}}(\mathbf{r},t)$ and $J_{\text{ind}}(\mathbf{r},t)$, respectively in the magnetic nanoprobe, leading to $f = \rho_{\text{ind}} E + J_{\text{ind}} \times B$. The total time-average force exerted on the probe is provided by volume integration, leading to

$$\mathbf{F}_{\text{tot}}^{\text{Lorentz}} = \tfrac{1}{2}\operatorname{Re}\left(\int_V \rho_{\text{ind}} \mathbf{E}^* + \mathbf{J}_{\text{ind}} \times \mathbf{B}^* dv\right) \tag{S2}$$

where non italic fonts denote phasors, * denotes complex conjugation, and we have implicitly assumed a $e^{-i\omega t}$ time dependence.

The volumetric integral (S2) and has been numerically calculated directly using COMSOL evaluated fields, to produce the results of $F_{tot,z}^{Lorentz}$ in Figs. 2, S3, and S4. The term $F_{B,z}^{Lorentz}$ in Fig. 2 has been calculated in the same way, but using only the term $\mathbf{J}_{ind} \times \mathbf{B}^*$ in (S2).

## **Optical force and the dipole approximation**

By neglecting all multipoles except the dipoles, one can model the nanoprobe as a dipolar scatterer with induced electric and magnetic dipole moments $\mathbf{p}$ and $\mathbf{m}$, respectively. The time-averaged optical force $\mathbf{F}$ exerted on the object for time harmonic fields with time dependence $e^{-i\omega t}$ is represented in terms of phasors as follows (*13*):

$$\mathbf{F}_{tot}^{dipole} = \frac{1}{2} \mathrm{Re} \left\{ \left( \nabla \mathbf{E}^{loc}(\mathbf{r}) \right)^* \cdot \mathbf{p} + \mu_0 \left( \nabla \mathbf{H}^{loc}(\mathbf{r}) \right)^* \cdot \mathbf{m} - \mu_0 \frac{ck^4}{6\pi} (\mathbf{p} \times \mathbf{m}^*) \right\}, \quad (S3)$$

The *i*-th component of the time-averaged optical force $F_i$ in Eq. (S3) reads,

$$\mathrm{F}_{tot,i}^{dipole} = \frac{1}{2} \mathrm{Re} \left\{ \Sigma_j \left[ \mathrm{p}_j \left( \partial_i \mathrm{E}_j^{loc} \right)^* + \mu_0 \mathrm{m}_j \left( \partial_i \mathrm{H}_j^{loc} \right)^* \right] - \mu_0 \frac{ck^4}{6\pi} (\mathbf{p} \times \mathbf{m}^*)_i \right\} \quad (S4)$$

for $i, j = x, y, z$ Cartesian coordinates. Note that one may consider the effect of higher order multipoles and calculate the optical force exerted on the object as in Ref. (*44*), however, within an acceptable approximation range it is enough to only include the effect of dipole moments in our analysis, as demonstrated in our results when comparing the dipolar force to the total Lorentz force.

## **The system of coupled Si probe-image under APB illumination**

Let us first demonstrate that the probe in the system of a coupled sphere under APB illumination is a magnetic scatterer at a specific wavelength range and under certain alignment condition. To that end, we calculate the multipoles (up to magnetic quadrupoles) of the top sphere (as the model of the probe) when it is coupled with the image sphere (that mimics the effect of the substrate). As the representation of the realistic fabricated probe, we design the probe and image spheres to have the comparable size with 92nm radius, and their magnetic resonance wavelength is around 610 nm. We prove that under a specific alignment between the two spheres and the beam, the probe sphere is exclusively a magnetic dipole scatterer. Indeed, we calculate and compare the power scattered by each multipole and show that under such an alignment the scattered power due to the magnetic dipole of the probe sphere is dominant. Accordingly, we consider only the first four multipoles, i.e., electric and magnetic dipole as well as quadrupole moments $\mathbf{p}$, $\mathbf{m}$, $\bar{\bar{\mathbf{Q}}}^e$, and $\bar{\bar{\mathbf{Q}}}^m$, respectively, which are enough for our analysis proof due to the small size of the scatterer within the frequency band of interest (i.e., $2r \sim 0.3\lambda_r$, where $2r = 184$nm is the diameter of the Si sphere and $\lambda_r = 610$nm is the magnetic resonance wavelength for this radius). The electric and magnetic dipole moments $\mathbf{p}$ and $\mathbf{m}$ are two vectors whereas the electric and magnetic quadrupole moments $\bar{\bar{\mathbf{Q}}}^e$ and $\bar{\bar{\mathbf{Q}}}^m$ are two tensors of second rank. The components of each moment read (*43, 45, 46*):

$$p_\alpha = \int_V r_\alpha \rho(\mathbf{r})\,dv, \qquad m_\alpha = \frac{1}{2}\int_V \left[\mathbf{r}\times\mathbf{J}(\mathbf{r})\right]_\alpha dv,$$
$$Q^e_{\alpha\beta} = \int_V \left(3r_\alpha r_\beta - r^2 \delta_{\alpha\beta}\right)\rho(\mathbf{r})\,dv, \qquad Q^m_{\alpha\beta} = \frac{1}{3}\int_V \left(\left[\mathbf{r}\times\mathbf{J}(\mathbf{r})\right]_\alpha r_\beta + \left[\mathbf{r}\times\mathbf{J}(\mathbf{r})\right]_\beta r_\alpha\right)dv, \tag{S5}$$

respectively. Here, in Cartesian coordinates, $\alpha$ and $\beta$ are $x$, $y$ and $z$ coordinates and such indices are used to represent Cartesian components of vectors and tensors, whereas $\mathbf{r} = x\hat{\mathbf{x}} + y\hat{\mathbf{y}} + z\hat{\mathbf{z}}$ is the position vector and $r = \sqrt{x^2 + y^2 + z^2}$. Moreover, $\delta_{\alpha\beta}$ is the Kronecker delta, $\rho$ and $\mathbf{J}$ are, respectively, the charge and current density distributions over the scatterer volume $V$, and the integrations are taken over the entire volume of the scatterer. It can be shown that the scattered power $P_{\text{scat}}$ due to these multipoles reads (*43*, *47*)

$$P_{\text{scat}} = \frac{\omega k^3}{12\pi}\left[\sum_\alpha \left(\frac{|p_\alpha|^2}{\epsilon_0} + \mu_0|m_\alpha|^2\right) + \frac{1}{120}\sum_{\alpha,\beta}\left(\frac{|kQ^e_{\alpha\beta}|^2}{\epsilon_0} + \mu_0|kQ^m_{\alpha\beta}|^2\right)\right]. \tag{S6}$$

Here $\omega$ and $k$ are angular frequency and wavenumber whereas $\epsilon_0$ and $\mu_0$ are the free space permittivity and permeability, respectively. Fig. S2 demonstrates the contribution to the total scattered power, Eq. (S6), generated by the Si probe sphere, provided by the electric and magnetic dipoles, and by the electric and magnetic quadrupoles defined in Eq. (S5). Results are obtained using full-wave numerical calculations carried out with COMSOL Multiphysics software (*34*) which is based on the finite element method (*33*) (FEM).

In Fig. S2 we plot the multipole contributions to the total scattered power given by Eq. (S6) for the probe sphere by moving the system of two spheres from the center of the beam axis (at $x = 0$ nm) toward the edge of the beam waist ($x = 450$ nm). The two spheres have a radius of 92nm radius nm, and the gap between them is 5 nm. In these calculations, we have used an APB with beam waist parameter $w_0 = 0.7\lambda$ and carrying a power of 150 µW coming from the bottom, with its minimum waist at $z = 0$. As it is clear from this figure, the scattering power due to the magnetic dipole is dominant when the system of two spheres is fully aligned with the axis of the excitation beam (at $x=0$). In such a scenario, the total scattered power peaks at $\lambda = 610$ nm which corresponds to its magnetic resonance. Interestingly, the footprint of the magnetic dipole always presents even when the nanoprobe is laterally displaced with respect to the beam axis when the scattering of the electric dipole is dominant.

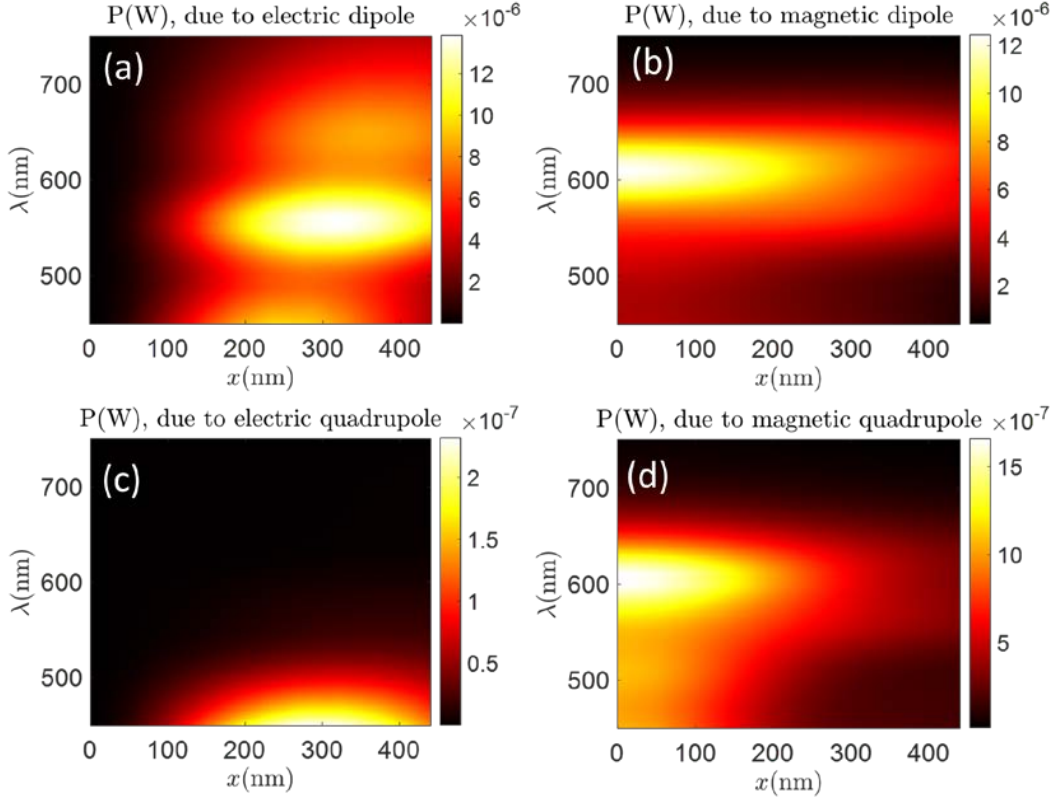

**Fig. S2.** Contribution of dominant multipole moments to the total scattered power generated by the top Si sphere, Eq. (S6), versus displacement $x$ between the position of the spherical nanoprobe and the APB axis. Position $x = 0$ denotes the case when the tip and sample spheres are fully aligned with the beam axis. (a) Electric dipole; (b) magnetic dipole; (c) electric quadrupole; and (d) magnetic quadrupole of the tip in the system of two coupled spheres.

### The exerted force on the Si probe sphere in the presence of its image sphere under the APB illumination

We have already proven in the previous section that scattering by the probe Si sphere at the presence of the image is dominated by the scattering of the magnetic dipole moment at the wavelength of interest, when illuminated by an APB. Here we calculate the exerted force on this magnetic Mie resonator integrating the Lorentz force density formulation of Eq. (S1), and demonstrate that the total force calculated using this force formulation on this magnetic resonator with a good approximation is equivalent to

$$F_{m,i}^{dipole} = \frac{1}{2}\mu_o \, \text{Re}\left\{\sum_j m_j \left(\partial_i H_j^{loc}\right)^*\right\}, \quad (S7)$$

where, $\partial_i$ is partial derivative with respect to the $i$-th spatial coordinate, and $i,j=x,y,z$ in Cartesian coordinates. This corresponds to the components of the magnetic dipole force expression in Eq. (3). Indeed, in the main body of the paper, this force was called as the magnetic dipolar force since it only includes the interaction between the magnetic dipole moment and the magnetic field gradient. In this paper, the magnetic dipole force in Eqs. (S7) and (3) is evaluated by the product of the numerically calculated magnetic dipole as in Eq. (S5) and the numerically calculated magnetic field gradient at the center of the tip sphere, by using the finite element method simulations implemented in COMSOL Multiphysics. When this formula is applied to the Si truncated cone tip, the gradient is evaluated at the center of the cone.

Figure S3 shows the total optical force map, calculated using (S2), exerted on the top Si *sphere* (representing the tip of the probe) and its image sphere illuminated by an APB from the bottom. The two spheres system is the same as the one considered in the previous section, i.e., the have a radius of 92nm radius nm, and the gap between them is 5 nm. The force is calculated at each position when we laterally move the Si tip and image spheres, relative to the APB illumination axis for two wavelengths: 610 nm (magnetic resonance) and 550 nm (electric resonance). As shown in this Figure, the force map shows a bright center circular shape with a hot spot at the magnetic resonance of the scatterer at 610 nm, and a doughnut shape at the electric resonance (550 nm) as it is also the case in the experimental measurements discussed in the main body of the manuscript. Importantly, the force map at the magnetic resonance resembles the magnetic field distribution as discussed in the main manuscript and will be shown in a next section here. Indeed, this Figure proves that the magnetic force and the magnetic field are proportional as shown in Eq. (4) in the manuscript and is proven below.

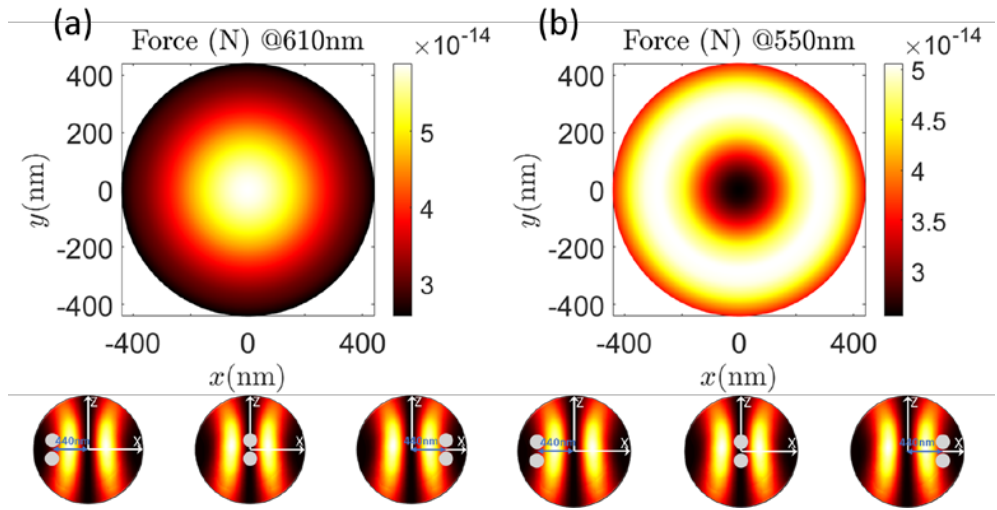

**Fig. S3**. Total optical force distribution exerted on the Si spherical probe in the system of probe-image spheres and APB illumination from the bottom, when the beam axis and the system of two spheres are relatively moved with respect to each other at (a) 610 nm and (b) 550 nm. The force is calculated using Eq. (S2) for an illumination power of 150 µW.

In Fig. S4 we compare the exerted force calculated from the Lorentz total force formulation in Eq. (S1) integrated over the whole top sphere with the contributions due to the electric and magnetic dipolar forces calculated from the dipole approximation formulation in Eq. (S3). The comparison is done at two wavelengths: at 610 nm, as the *on-state* condition (on magnetic resonance), and at 550 nm, as the *off-state* condition (off magnetic resonance).

The different terms in Fig. S4 are all longitudinal force components defined as follows: the time-averaged *total* optical force $F_{tot,z}^{Lorentz}$ exerted on the nanoprobe is defined in Eq. (S2); the time-average *magnetic dipolar force* $F_{m,z}^{dipole}$ is defined in Eqs. (3) and (S7); the time-average second term in the Lorentz formula due to the magnetic field **B** is $F_{B,z}^{Lorentz} = \frac{1}{2}\text{Re}\left(\int_V \mathbf{J}_{ind} \times \mathbf{B}^* dv\right)_z$; and the time-average electric dipolar force $F_{p,z}^{dipole}$ is defined in Eqs. (2). As shown in Fig. S4(a), the total force $F_{tot,z}^{Lorentz}$ exerted on the tip is in close agreement with the magnetic dipolar force $F_{m,z}^{dipole}$ in a wide range of relative positions between the beam axis and the two coupled scatterers at the magnetic resonance wavelength whereas the agreement does not hold for the off-magnetic resonant wavelength shown in Fig. S4(b). This proves that the total optical force is purely magnetic within the interested wavelength range and under this specific type of excitation scenario.

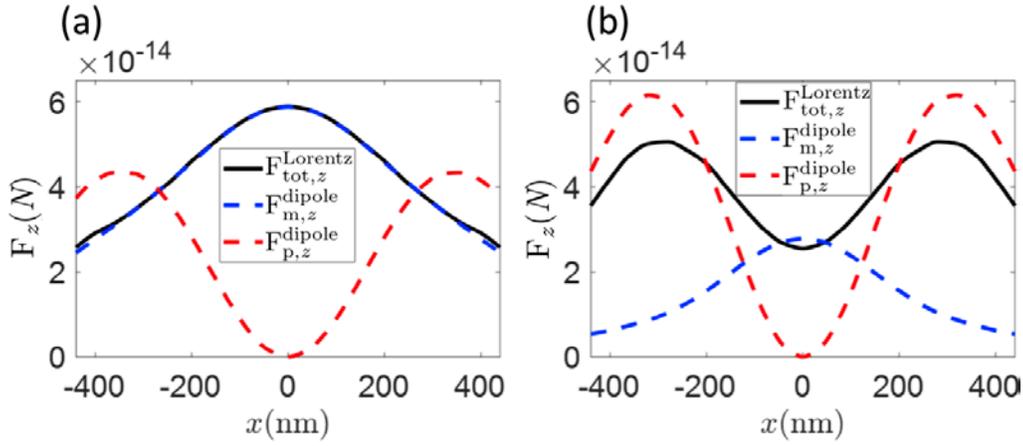

**Fig. S4.** Comparison of optical force exerted on the Si *spherical* tip in the system made of tip-image spheres under APB illumination, at the magnetic resonance 610 nm (a), and off-resonance 550 nm (b). The force $F_{tot,z}^{Lorentz}$ is numerically calculated using Eq. (S2). The dipolar forces $F_{p,z}^{dipole}$ and $F_{m,z}^{dipole}$ are respectively evaluated using Eqs. (2) and (3) in the main body of the paper. The *x* coordinate represents the displacement of the two spheres with respect to the beam axis. When on-resonance, the accurate Lorentz force $F_{tot,z}^{Lorentz}$ is well represented by the magnetic dipole force $F_{m,z}^{dipole}$.

**<u>Derivation of Eq. (4) of the main manuscript</u>**

Referring to Eq. (S7), the local magnetic field, $H_j^{loc}$ possesses one contribution from the incident APB beam and the other one from the scattered beam due the reflections from the substrate. Here by using the image principle (*43*), we replace the substrate with an image dipole, and consider its interactive effect in the field contributions from the image dipole. Therefore, the local magnetic field at the tip position reads

$$\mathbf{H}^{loc}(\mathbf{r}_{tip}) = \mathbf{H}^{inc}(\mathbf{r}_{tip}) + \mathbf{H}^{scat}_{img \to tip}(\mathbf{r}_{tip}) \tag{S8}$$

where,

$$\mathbf{H}^{scat}_{img \to tip}(\mathbf{r}_{tip}) = \underline{\mathbf{G}}(\mathbf{r}_{tip} - \mathbf{r}_{img}) \cdot \mathbf{m}_{img} \tag{S9}$$

and $\underline{\mathbf{G}}(\mathbf{r}) = G_x \hat{\mathbf{x}}\hat{\mathbf{x}} + G_y \hat{\mathbf{y}}\hat{\mathbf{y}} + G_z \hat{\mathbf{z}}\hat{\mathbf{z}}$ is the magnetic dyadic Green's function in Cartesian coordinates with $\mathbf{r}$ being the position vector at the tip/image location. Due to optically small distance between the tip and its image, the near field term in the Green's function is dominant and is approximated by (*43*):

$$\underline{\mathbf{G}}(\mathbf{r}) = \frac{1}{4\pi |\mathbf{r}|^3}(3\hat{\mathbf{r}}\hat{\mathbf{r}} - \underline{\mathbf{I}}) \tag{S10}$$

where $\underline{\mathbf{I}}$ is the identity tensor of rank 2 and $\hat{\mathbf{r}}$ is the unit vector from the source to the observation point. The dipole moments of the image dipole and the tip dipole are

$$\mathbf{m}_{img} = \underline{\boldsymbol{\alpha}}^m_{img} \cdot \mathbf{H}^{loc}(\mathbf{r}_{img}), \tag{S11}$$

$$\mathbf{m}_{tip} = \underline{\boldsymbol{\alpha}}^m_{tip} \cdot \mathbf{H}^{loc}(\mathbf{r}_{tip}). \tag{S12}$$

Here $\underline{\boldsymbol{\alpha}}^m_{tip}$ and $\underline{\boldsymbol{\alpha}}^m_{img}$ are the dipolar magnetic polarizability tensors of the tip and image, respectively. Similarly, the local magnetic field at the image position reads

$$\mathbf{H}^{loc}(\mathbf{r}_{img}) = \mathbf{H}^{inc}(\mathbf{r}_{img}) + \mathbf{H}^{scat}_{tip \to img}(\mathbf{r}_{img}) \tag{S13}$$

where,

$$\mathbf{H}^{scat}_{tip \to img}(\mathbf{r}_{img}) = \underline{\mathbf{G}}(\mathbf{r}_{img} - \mathbf{r}_{tip}) \cdot \mathbf{m}_{tip} \tag{S14}$$

Inserting Eq. (S14) in Eq. (S13) and then by using the resulting equation in Eq. (S11), we derive a closed-form expression for $\mathbf{m}_{img}$ which is eventually used to obtain $\mathbf{H}^{loc}(\mathbf{r}_{tip})$ from Eqs. (S8) and (S9). Therefore, the longitudinal component of the optical force at the tip position from Eq. (S7) reads

$$F^{dipole}_{m,z} = \frac{1}{2}\mu_o \operatorname{Re}\left\{ m_{tip,x}\left(\partial_z H^{loc}_x(\mathbf{r}_{tip})\right)^* + m_{tip,y}\left(\partial_z H^{loc}_y(\mathbf{r}_{tip})\right)^* + m_{tip,z}\left(\partial_z H^{loc}_z(\mathbf{r}_{tip})\right)^* \right\} \tag{S15}$$

Finally, by using Eq. (S12) in the above equation, we obtain

$$F^{dipole}_{m,z} = \frac{3}{8\pi d^4} \operatorname{Re} \sum_j \Big\{ \left(\alpha^m_{tip,xj} H^{inc}_j(\mathbf{r}_{tip})\right)\left(\alpha^m_{img,xj} H^{inc}_j(\mathbf{r}_{img})\right)^* + \left(\alpha^m_{tip,yj} H^{inc}_j(\mathbf{r}_{tip})\right)\left(\alpha^m_{img,yj} H^{inc}_j(\mathbf{r}_{img})\right)^*$$
$$-2\left(\alpha^m_{tip,zj} H^{inc}_j(\mathbf{r}_{tip})\right)\left(\alpha^m_{img,zj} H^{inc}_j(\mathbf{r}_{img})\right)^* \Big\} \tag{S16}$$

where, $\alpha^m_{tip,ij}$ and $\alpha^m_{img,ij}$ ($i, j = x, y, z$) are the magnetic polarizability components of the tip and the image dipole, respectively, in Cartesian coordinates and $d$ is the vertical distance between the tip and the image dipole. In obtaining Eq. (S16), we considered that the tip and its image are positioned along the *z*-axis. Moreover, in determining the local magnetic fields, we neglected terms containing incident fields compared to terms containing the gradient of scattered fields since their values are much smaller in the near zone of a scatterer. Besides, we have ignored all the terms containing polarizability power orders higher than second and since the dipole and its image are optically very close, we have neglected the terms containing $G = 1/\left(4\pi |z|^3\right)$ compared to those containing $\partial G/\partial z = -3/\left(4\pi |z|^4\right)$.

Next, if the phase difference of the incident beam between the tip and its image is neglected due to their deep subwavelength distance, we can assume that $\mathbf{H}^{inc}(\mathbf{r}_{tip}) \approx \mathbf{H}^{inc}(\mathbf{r}_{img}) = \mathbf{H}^{inc}$ and the force reduces to

$$F_{m,z}^{dipole} = \frac{3}{8\pi d^4} \operatorname{Re} \sum_j \left\{ \left(\alpha_{tip,xj}^m H_j^{inc}\right)\left(\alpha_{img,xj}^m H_j^{inc}\right)^* + \left(\alpha_{tip,yj}^m H_j^{inc}\right)\left(\alpha_{img,yj}^m H_j^{inc}\right)^* \right. \\ \left. -2\left(\alpha_{tip,zj}^m H_j^{inc}\right)\left(\alpha_{img,zj}^m H_j^{inc}\right)^* \right\}$$

(S17)

If one considers a reference system such that the polarizability tensors are diagonal, i.e., $\alpha_{img,ij}^m = \alpha_{tip,ij}^m = 0; \ i \neq j$, and also assume an azimuthally symmetric scatterer, i.e., $\alpha_{tip,xx}^m = \alpha_{tip,yy}^m = \alpha_{tip,zz}^m = \alpha_{tip}^m$ and $\alpha_{img,xx}^m = \alpha_{img,yy}^m = \alpha_{img,zz}^m = \alpha_{img}^m$, the time-averaged optical force reads

$$F_{m,z}^{dipole} = \frac{3}{8\pi d^4} \operatorname{Re} \left\{ \left(\alpha_{tip}^m H_x^{inc}\right)\left(\alpha_{img}^m H_x^{inc}\right)^* + \left(\alpha_{tip}^m H_y^{inc}\right)\left(\alpha_{img}^m H_y^{inc}\right)^* \\ -2\left(\alpha_{tip}^m H_z^{inc}\right)\left(\alpha_{img}^m H_z^{inc}\right)^* \right\}$$

(S18)

For the case of Azimuthally polarized beam we have only z component of magnetic field at the tip position, $H_x^{inc} = 0$ and $H_y^{inc} = 0$. Also, since the polarizability of the image dipole is proportional to the tip dipole $\alpha_{tip}^m \propto \alpha_{img}^m$, from Eq. (S18) the time-averaged optical force on the tip is related to the incident magnetic field at the tip position as $F_z \propto \left|\alpha_{tip}^m H_z^{inc}\right|^2$ which is Eq. (4) of the manuscript. Note that we have suppressed superscript "dipole" and subscript "m" in the notation of this section for simplicity.

### Approximation condition of Eq. (4) with axis displacement for the real scenario with a truncated Si cone

Lastly, we analyze the approximation condition of Eq. (4) in the main paper, i.e., the proportionality between the *magnetic force* and the *longitudinal magnetic field intensity*, even when the probe is slightly misaligned with respect to the axis of the incident APB. For this purpose, we define the *normalized longitudinal magnetic* field intensity to the maximum magnetic force as

$$\left|H_z^{inc}\right|^2 \frac{\text{Max}(F_{m,z})}{\text{Max}\left(\left|H_z^{inc}\right|^2\right)}.$$

We perform the simulation of the realistic scenario as indicated in the main paper: the on-state truncated cone probe above the glass substrate is scanning in the transverse direction (*x*) around the axis of the incident APB, at 670 nm wavelength. The APB has a beam waist parameter of $w_0 = 0.7\lambda$ with incident beam power of 150 μW. We compare the normalized longitudinal magnetic field intensity and the longitudinal magnetic force with respect to the displacement *x* between the axis of the incident APB and the on-state probe, shown in Fig. S5. The light blue curve is the force due to induced magnetic dipole in the on-state truncated Si disk calculated by Eq. (5) whereas the dark blue curve is the longitudinal incident magnetic field intensity normalized to maximum force due to induced magnetic dipole.

We highlight the yellow region where the normalized longitudinal magnetic field and the magnetic force have an overlap accuracy (ratio) higher than 90%: the spot has a remarkable diameter close to 400 nm. Considering the wavelength for on-state magnetic excitation as 670 nm, the highlighted region shows a rather large area near APB axis where the proportionality between the longitudinal magnetic force and the incident magnetic field intensity holds.

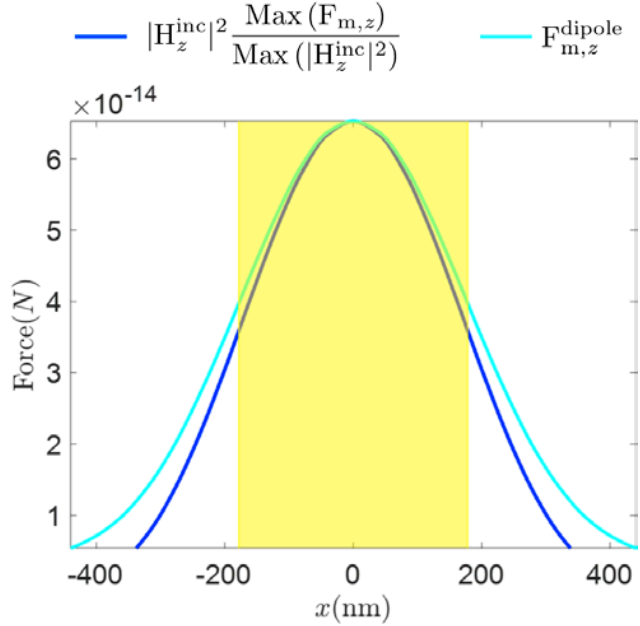

**Fig. S5.** Comparison of the normalized longitudinal magnetic incident field intensity and longitudinal magnetic force exerted on the magnetic nanoprobe (on-state Si truncated cone) 5nm over the dielectric substrate versus displacement *x* with respect to the axis of the incident APB. The yellow highlight indicates the region where the ratio between the normalized longitudinal magnetic field intensity and the magnetic force is greater than 90%. The magnetic dipole force $F_{m,z}^{dipole}$ is calculated using Eq. (3).

**Fabrication, optical characterization and force measurement**

The Si truncated cone probe is made based on the commercial Si atomic force microscopy probe from Nanosensors (PPP-NCHR), which has the nominal mechanical resonance frequency at 330 kHz. We use the focused ion beam (FIB) lithography to drill hole from the side of the Si probe and remove its sharp tip by using the Quanta 3D FEG Dual Beam scanning electron microscope (SEM) from FEI Thermo Fisher Scientific. The Si material of the commercial probe is single crystalline silicon doped with Antimony (N-doped). According to the vendor, the permittivity of this Si is well characterized and used based on the previous literature (*48*).

We use the continuous wave (CW) single mode fiber pigtailed laser diode with 670 nm wavelength from Thorlabs Inc. as the light source. We use the polarization converter from ARCoptix to generate the APB.

We perform all the photoinduced force measurement by using the commercial PiFM Vistascope from Molecular Vista Inc. Inside the Vistascope, the incident APB is sharply focused by a high numerical aperture oil-immersed objective lens from Olympus. We optimize various parameters/settings of the Vistascope to improve SNR measurement results. The minimum detectable force of the Vistascope can reach sub-pico Newtons range (*49*).